\documentclass[aip,jmp,reprint,amsmath,amssymb,onecolumn,numerical]{revtex4-1}
\usepackage[utf8]{inputenc}
\usepackage{graphicx,dsfont,xcolor}
\usepackage{cancel}
\usepackage{bbold}
\usepackage{titlesec}

\makeatletter
\g@addto@macro\bfseries{\boldmath}
\makeatother

\usepackage{hyperref}

\begin{document}

\begin{abstract}
  \noindent Time reversal symmetry is a fundamental property of many quantum mechanical systems. The relation between statistical physics and time reversal is subtle and not all statistical theories conserve this particular symmetry, most notably hydrodynamic equations and kinetic equations such as the Boltzmann equation. In this article it is shown analytically that quantum kinetic generalizations of the Boltzmann equation that are derived using the nonequilibrium Green functions formalism as well as all approximations that stem from $\Phi$-derivable selfenergies are time reversal invariant. 
  
\end{abstract}
\title{Time Reversal Invariance of quantum kinetic equations: Nonequilibrium Green Functions Formalism}
\author{Miriam Scharnke, Niclas Schlünzen, and Michael Bonitz}
\maketitle
\date{\today}

\renewcommand{\i}{\mathrm{i}}
\renewcommand{\d}{\mathrm{d}}
\renewcommand{\H}{\mathrm{H}}
\renewcommand{\S}{\mathrm{S}}

\section{Introduction}
\noindent
The nonequilibrium Green functions (NEGF) formalism provides an ab initio description of strongly interacting quantum many particle systems far from equilibrium. 
It has gained much importance in the last two decades, mainly because it is now possible to solve the two-time Keldysh--Kadanoff--Baym equations (KBE) numerically. 
NEGFs have been successfully used to describe a huge variety of systems and phenomena, such as Bose condensation, quantum and molecular transport~\cite{uimonen_comparative_2011} and femtosecond spectroscopy, carrier dynamics in quantum dots and quantum wells~\cite{gartner_relaxation_2006,lorke_influence_2006}, laser exciation of small atoms \cite{dahlen_solving_2007, balzer_efficient_2010}, nuclear collisions \cite{rios_towards_2011}, intense laser-plasma interaction \cite{bonitz_quantum_1999}, baryogenesis in cosmology\cite{garny_leptogenesis_2013} and much more. 
Within the Green functions formalism there exists an elegant diagrammatic method for constructing approximations that conserve energy, momentum, angular momentum and particle number, by using so-called 
$\Phi$-derivable selfenergies. It is the purpose of this paper to show that those approximations as well as the exact equations of motion of the Green functions formalism are invariant under 
time reversal.

The relation between time reversal symmetry and statistical physics is subtle and not all statistical theories are invariant under time reversal, the most famous counterexample being the 
Boltzmann equation of classical statistical mechanics and its quantum generalization. Therefore, extensive work has been done over the recent seven decades to derive non-Markovian generalizations of the Boltzmann equation that are time-reversal invariant as the underlying quantum mechanical system. Among the well established approaches we mention density operator concepts, see e.g. [\onlinecite{bonitz_qkt}] for an overview, and nonequilibrium Green functions [\onlinecite{kadanoff-baym-book}], for a recent text book discussion, see [\onlinecite{stefanucci}]. Despite recent activities in this field we are not aware of a general analysis of the time reversal properties of the resulting generalized quantum kinetic equations. Since these equations are usually solved with the help of certain many-body approximations, it is even more important to understand under which conditions time reversal invariance is retained.
%

It is the goal of the present article to solve these questions for the NEGF formalism which we briefly recall in Sec.~\ref{NEGFs}. Since the Kadanoff--Baym equations can be directly derived from the  equations of motion of the field operators in second quantization which are time-reversal invariant, it should be expected that the KBE  have the same symmetry properties. 
It is, nonetheless, not trivial to show this directly in full generality, and a successful procedure is presented in Sec.~\ref{s:tri-msh}. We then demonstrate in Sec.~\ref{s:tri-phi} that an important class of approximations---the so-called $\Phi$-derivable approximations---also preserve time reversal symmetry. We conclude with a summary in Sec.~\ref{s:summary} where we also outline the time reversal invariance conditions of the generalized Kadanoff-Baym ansatz \cite{lipavsky}.

\section{Nonequilibrium Green Functions}
\label{NEGFs}
\begin{figure}
 \includegraphics[width=0.5\linewidth]{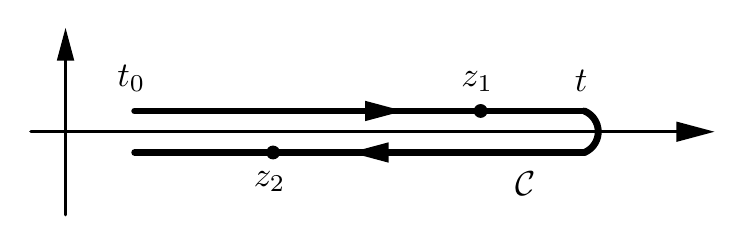}
 \caption{Illustration of the two real-time branches of the Keldysh contour. $z_1$ on the causal branch $\mathcal{C}_-$ is earlier on the contour than $z_2$ on the anti-causal branch 
 $\mathcal{C}_+$, although the physical time $t_1$ corresponding to $z_1$ is later than the physical time $t_2$ corresponding to $z_2$. 
 }
 \label{fig:keldysh_contour}
\end{figure}

\noindent The $n$-particle Green function $\mathrm{G}^{(n)}$ is defined element-wise as the ensemble average of the $n$-particle correlator in second quantization
\begin{align}
   &G_{i_1 \dots i_n;j_1 \dots j_n}^{(n)}(z_1 \dots z_n; z_1' \dots z_n') = \left\langle\hat{G}_{i_1 \dots i_n;j_1 \dots j_n}^{(n)}(z_1 \dots z_n; z_1' \dots z_n') \right\rangle\nonumber\\
  &=\left(-\frac{\i}{\hbar}\right)^n \left\langle \hat{\mathcal{T}}_{\mathcal{C}} \: \hat{c}_{i_1}(z_1) \dots \hat{c}_{i_n}(z_n)\hat{c}^{\dagger}_{j_n}(z_n')\dots \hat{c}^{\dagger}_{j_1}(z_1')\right\rangle\;,
\end{align}
where $\hat{c}^{\dagger}_{j_k}$ and $\hat{c}_{i_k}$ are second quantization creation and annihilation operators with respect to a complete orthonormal basis of single-particle states $\{ |\phi_i\rangle\}$ obeying the (anti-)commutation relations for bosons (fermions)
\begin{align}
[\hat{c}_{i_k}, \hat{c}_{i_l}]_{\mp} & = [\hat{c}^{\dagger}_{i_k}, \hat{c}^{\dagger}_{i_l}]_{\mp} = 0,
\nonumber\\
[\hat{c}_{i_k}, \hat{c}^\dagger_{i_l}]_{\mp} & = \delta_{i_k, i_l}.
\end{align}
Further, 
$\hat{\mathcal{T}}_{\mathcal{C}}$ is the time ordering operator on the Keldysh 
time contour $\mathcal{C}$, as illustrated in Fig. \ref{fig:keldysh_contour}.

\noindent
The dynamics of the $n$-particle Green function are described by the Martin--Schwinger hierarchy--a coupled hierarchy of equations of motion (we leave out the orbital indices for brevity):
\begin{align}
\label{eq:correlator_eom1}
 &\left[ \i\hbar\partial_{z_k} - h^{(0)}(z_k)\right]\mathrm{G}^{(n)}(z_1 \dots z_n;z_1' \dots z_n') = 
 \\
 &\pm\i\hbar\int_{\mathcal{C}}\d\bar{z}\,W(z_k\bar{z})\,\mathrm{G}^{(n+1)}(z_1 \dots z_n\bar{z};z_1' \dots z_n'\bar{z}^+)
 +\sum_{p=1}^n(\pm1)^{k+p}\delta_{\mathcal{C}}(z_kz_p')\mathrm{G}^{(n-1)}(z_1 \dots \bcancel{z_k} \dots z_n;z_1' \dots \bcancel{z_p'} \dots z_n'),
\nonumber
\end{align}
and
\begin{align}
\label{eq:correlator_eom2}
&\mathrm{G}^{(n)}(z_1 \dots z_n;z_1' \dots z_n')\left[-\i\hbar\overleftarrow{\partial}_{z_k'} - h^{(0)}(z_k')\right] = 
\\
 &\pm\i\hbar\int_{\mathcal{C}}\d\bar{z}\,\mathrm{G}^{(n+1)}(z_1 \dots z_n\bar{z}^-;z_1' \dots z_n'\bar{z})W(\bar{z}z_k')
 +\sum_{p=1}^n(\pm1)^{k+p}\delta_{\mathcal{C}}(z_pz_k')\mathrm{G}^{(n-1)}(z_1 \dots \bcancel{z_p} \dots z_n;z_1' \dots \bcancel{z_k'} \dots z_n') \;,
\nonumber 
\end{align}
where $W(z_1, z_2) = \delta_{\mathcal{C}}(z_1, z_2)w(z_1)$ and $w(z_1)$ is the instantaneous two-particle interaction operator. 
The first-order hierarchy equations can be formally closed by introducing  the selfenergy $\Sigma$, reducing the description   to the dynamics of the single-particle Green function $\mathrm{G}^{(1)}$:
 \begin{equation}
\label{eq:KBE1}
  \left[\i\hbar\partial_z - h(z)\right]\mathrm{G}^{(1)}(zz') = \delta_{\mathcal{C}}(zz')\,\mathbb{1} + \int_{\mathcal{C}} \d\bar{z}\, \Sigma(z\bar{z})\, \mathrm{G}^{(1)}(\bar{z}z')\;,
 \end{equation}
and its adjoint, 
\begin{equation}
\label{eq:KBE2}
 \left[-\i\hbar\partial_{z'} - h(z)\right]\mathrm{G}^{(1)}(zz') = \delta_{\mathcal{C}}(zz')\mathbb{1} + \int_{\mathcal{C}} \d\bar{z}\, \mathrm{G}^{(1)}(z\bar{z})\Sigma(\bar{z}z')\;.
\end{equation}
These equations are the Keldysh--Kadanoff--Baym equations. It is theoretically possible to write $\Sigma$ as a functional of $\mathrm{G}^{(1)}$ such that Eqs. \eqref{eq:KBE1} and \eqref{eq:KBE2} are still exact. 
The main challenge of the Green functions formalism is to find suitable approximations for the selfenergy. One important class of selfenergy approximations is constructed as 
the functional derivative of a scalar function $\Phi$ (``$\Phi$-derivable approximations''):
\begin{equation}
\label{eq:sigma_phi}
 \Sigma(z_1,z_2) = \dfrac{\delta\Phi[G]}{ \delta G(z_2, z_1^+)}.
\end{equation}
These approximations conserve particle number, momentum, energy, and angular momentum, if $\Phi$ is invariant under gauge transformations, space and time translations, and rotations, respectively. 
This is satisfied if $\Phi$ is the amplitude of a scattering process (since every scattering process satisfies these conservation requirements). Therefore, it is possible to construct conserving 
scalar potentials diagrammatically.
\begin{figure}
 \includegraphics[width=.5\linewidth]{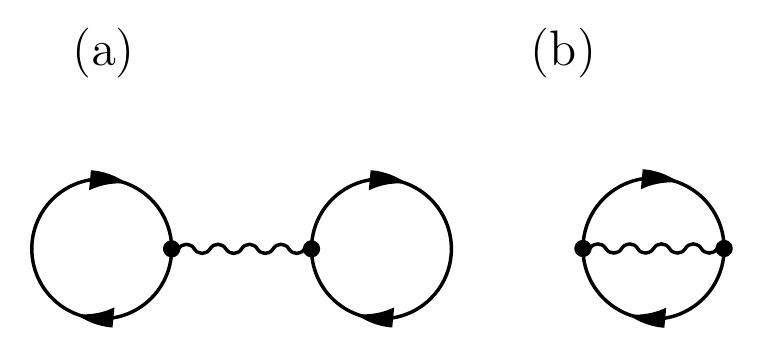}
 \caption{The Hartree (a) and Fock (b) diagrams contributing to the Hartree--Fock potential $\Phi^{\mathrm{HF}}$. A full (wiggly) line corresponds to a Green function (interaction potential).}
 \label{fig:phi_hf_diag}
\end{figure}
For example, the potential $\Phi^{\mathrm{HF}}$ corresponding to the Hartree--Fock approximation consists of two diagrams, as illustrated in Fig. \ref{fig:phi_hf_diag}:
\begin{align}
  \Phi^{\mathrm{HF}}[\mathrm{G}] &= \pm\dfrac{\i\hbar}{2} \Phi^{(\mathrm{a})}[\mathrm{G}] + \dfrac{\i\hbar}{2} \Phi^{(\mathrm{b})}[\mathrm{G}]\;,
  \label{eq:phi_hf}\\
\Phi^{(\mathrm{a})}[\mathrm{G}] &= \int_{\mathcal{C}}\d z_1\d z_2 \;\mathrm{G}(z_1;z_1^+)W(z_1;z_2)\mathrm{G}(z_2;z_2^+)\;,\\
 \Phi^{(\mathrm{b})}[\mathrm{G}] &= \int_{\mathcal{C}}\d z_1\d z_2 \;\mathrm{G}(z_1;z_2^+)W(z_1;z_2)\mathrm{G}(z_2;z_1^+)\;,
\end{align}
resulting in the Hartree--Fock selfenergy
\begin{align}
\label{eq:sigma_hf}
 \Sigma^{\mathrm{HF}}(zz') =& \pm\i\hbar \delta_{\mathcal{C}}(zz')\int_{\mathcal{C}}\d\bar{z}\,W(z\bar{z})\mathrm{G}^{(1)}(\bar{z}\bar{z}^+) 
 + \i\hbar \mathrm{G}^{(1)}(zz')W(z^+z')\;.
\end{align}
Other $\Phi$-derivable approximations are the second order and third order Born approximations, the GW-approximation and the T-matrix (ladder) approximation, cf. e.g. Ref.~[\onlinecite{stefanucci}].
\section{Time Reversal Invariance in Quantum Many-Body Theory}
Here, we briefly recall the notion of time reversibility inctroducing the time reversal operator ${\hat T}$. We first illustrate this for the $N$-particle Schrödinger equation and then extend the concept to many-body theory within second quantization.
\subsection{Time Reversal Invariance of the Schr\"odinger Equation}\label{sec:tri_se}
\noindent The Schr\"odinger equation is called symmetric with regard to time reversal if, (i) for any solution $|\psi(t)\rangle$, there exists another solution $|\psi'(t')\rangle$ with $t' = -t$, and if (ii) there exists a unique relation between the two: $|\psi'\rangle = \hat{T}|\psi\rangle$, for some operator $\hat{T}$\cite{haake}. It can be shown that $\hat{T}$ must not only 
be a linear operator, but an anti-unitary one. Thus, it can be expressed as the product of complex conjugation and some unitary operator $\hat{U}$. 
The quantum mechanical equivalent to classical conventional time reversal is obtained by choosing $\hat{U} = \mathds{1}$, so that 
$|\psi\rangle\rightarrow|\psi\rangle^*$.\\
Let us illustrate this for the time-dependent Schr\"odinger equation:
\begin{equation}
\label{eq:tdTSE}
 \i\hbar\,\partial_t|\psi\rangle = \hat{H}|\psi\rangle\;.
\end{equation}
Applying $\hat{T}$ to both sides yields:
\begin{align}
 \hat{T}\,\i\hbar\,\partial_t|\psi\rangle &= \hat{T}\,\hat{H}|\psi\rangle\nonumber\\
 \Longleftrightarrow \; \underbrace{-\i\hbar\,\partial_t}_{\i\hbar\partial_{(-t)}}\hat{T}|\psi\rangle &= \hat{T}\hat{H}\hat{T}^{-1}\hat{T}|\psi\rangle\;,
\end{align}
which means that $\hat{T}|\psi\rangle$ solves the time reversed Schr\"odinger equation 
\begin{equation}
 \i\hbar\, \partial_{(-t)}|\psi'\rangle = \hat{H}|\psi'\rangle
\end{equation}
if (and only if) $\hat{H}=\hat{T}\hat{H}\hat{T}^{-1}$. This result is valid for an arbitrary interacting many-particle system. 
\subsection[Heisenberg Equation]{Time Reversal Invariance of the Heisenberg Equation}
\label{Heisenberg}
The Heisenberg equation for an operator $\hat{A}_\H$ is equivalent to the Schr\"odinger equation and should, therefore, possess the same reversibility properties. This is
straightforwardly shown applying the ${\hat T}$-operator introduced above from left and right:
\begin{align}
 \i\hbar \,\partial_t\hat{A}_\H &= \left[ \hat{A}_{\H}, \hat{H}\right] \label{eq:HE}\\
 \Longleftrightarrow \; \hat{T}\,\i\hbar\,\partial_t\hat{A}_{\H}\, \hat{T}^{-1} &= \hat{T}\left( \hat{A}_{\H}\hat{H} - \hat{H}\hat{A}_{\H}\right)\hat{T}^{-1} \nonumber\\
 \Longleftrightarrow \; -\i\hbar\,\partial_t\hat{T}\hat{A}_{\H} \hat{T}^{-1} &= \hat{T}\hat{A}_{\H}\hat{T}^{-1}\,\hat{T}\hat{H}\hat{T}^{-1} - \hat{T}\hat{H}\hat{T}^{-1}\,\hat{T}\hat{A}_{\H}\hat{T}^{-1}\;,
\end{align}
which is equivalent to
\begin{equation}
\i\hbar\,\partial_{-t}\hat{T}\hat{A}_{\H} \hat{T}^{-1} = \left[\hat{T}\hat{A}_{\H}\hat{T}^{-1}, \hat{H}\right]
\end{equation}
if and only if $\hat{H} = \hat{T} \hat{H} \hat{T}^{-1}$. This means that, if a Heisenberg operator $\hat{A}_{\H}(t)$ solves the Heisenberg equation, then $\hat{T}\hat{A}_{\H}\hat{T}^{-1}$ solves the time-reversed 
Heisenberg equation. 
\subsection[Equations of Motion of the Field Operators of Second Quantization]{Time Reversal Invariance of the Equations of Motion of the Field Operators of Second Quantization}
\noindent
The equation of motion of the annihilation operator in an arbitrary single-particle basis $\{|\phi_i\rangle \}$ [cf. Sec.~\ref{NEGFs}] reads\cite{schluenzen2}
\begin{equation}
\label{eq:eom_anop}
 \i\hbar\,\partial_t\hat{c}_i(t) =  \sum_k \Big( t_{ik} + v_{ik}(t) \Big) \hat{c}_k(t) + \sum_{jkl}w_{ijkl}\,\hat{c}^{\dagger}_j(t) \hat{c}_l(t) \hat{c}_k(t)\;,
\end{equation}
where $t_{ik}$ and $v_{ik}$ ($w_{ijkl}$) are matrix elements computed with the respective single-particle (two-particle product) basis states.
For the purpose of analyzing time reversal symmetry, it is convenient to consider that Eq. \eqref{eq:eom_anop} is derived from and equivalent to the Heisenberg equation 
for $\hat{c}_i(t)$,
\begin{equation}
 \i\hbar\,\partial_t\hat{c}_i(t) = \left[ \hat{c}_i, \hat{H}\right],
\end{equation}
and, as such, possesses the same symmetry properties that $\hat{H}$ does. The same obviously holds for the creation operator $\hat{c}^{\dagger}_j$.

\section[Martin--Schwinger Hierarchy]{Time Reversal Invariance of the Martin--Schwinger Hierarchy}\label{s:tri-msh}
The Martin--Schwinger hierarchy (\ref{eq:correlator_eom1}) follows from taking the ensemble average of the formally equivalent hierarchy of equations of motion of the $n$-particle correlators $\hat{\mathrm{G}}^{(n)}$. The latter, in turn, follows from the equations of motion of the field operators \eqref{eq:correlator_eom1}, \eqref{eq:correlator_eom2} and, therefore, must satisfy the same symmetry properties as the field operators. Nonetheless 
it is instructive to prove the time reversal invariance of the Martin--Schwinger hierarchy directly.
To this end, it is important to understand how the contour-$\delta$-distribution behaves under time reversal. Since $\delta$ is even with respect to its argument, i.e., $\delta(z) = \delta(-z)$, it might be expected 
that $\tilde{\delta_{\mathcal{C}}} := \delta_{\mathcal{C}}^{z\rightarrow-z}=\delta_{\mathcal{C}}$. That, however, cannot be true, as the following considerations show:
\begin{align}
    \int_{\mathcal{C}} \d z \, \delta_{\mathcal{C}} (z) = 1  &\xrightarrow{z \rightarrow -z} 1 = \int_{\mathcal{C}} \d (-z)\, \tilde{\delta_{\mathcal{C}}}(z) \;,\nonumber\\
    \Longleftrightarrow \;\;\delta_{\mathcal{C}} &\xrightarrow{z \rightarrow -z}\tilde{\delta_{\mathcal{C}}} = -\delta_{\mathcal{C}}\;.\label{eq:delta_rev}
\end{align}
This means that the $\delta$-distribution with respect to contour time arguments changes its sign under time reversal, in analogy to differential and integral operators.

Component-wise, the $n$-th order hierarchy equations for the correlators read
\begin{align}
 &\sum_l\left[\i\hbar\dfrac{\partial}{\partial z_k}\delta_{i_kl}-h_{i_kl}^{(0)}(z_k)\right]\hat{G}^{(n)}_{i_1 \dots l \dots i_n;j_1 \dots j_n}(z_1 \dots z_n;z_1' \dots z_n')\nonumber\\
 &= \pm\i\hbar\sum_{lmn}\int_{\mathcal{C}}\d\bar{z}\;\Big\{ W_{i_klmn}(z_k\bar{z})
 \hat{G}^{(n+1)}_{i_1 \dots m \dots  i_nn;j_1 \dots j_nl}(z_1 \dots z_n\bar{z};z_1' \dots z_n'\bar{z}^+)\Big\}\nonumber\\
 &+\sum_{p=1}^n \Big\{(\pm1)^{k+p}\delta_{i_kj_p}\delta_{\mathrm{C}}(z_kz_p')
 \hat{G}^{(n-1)}_{i_1 \dots \bcancel{i_k} \dots i_n;j_1 \dots \bcancel{j_p} \dots  j_n}(z_1 \dots \bcancel{z_k} \dots z_n;z_1' \dots \bcancel{z_p'} \dots z_n')\Big\}\;.\label{eq:corr_element}
\end{align}
Since $W(zz') = w(z)\delta_{\mathcal{C}}(zz')$, it immediately follows that 
$ W^{z^{(\prime)}\rightarrow-z^{(\prime)}} = - W$\;.
Therefore, the time-reversed equations read
\begin{align}
 &\sum_l\left[-\i\hbar\dfrac{\partial}{\partial z_k}\delta_{i_kl}-h_{i_kl}^{(0)}(z_k)\right]\hat{G}^{(n)}_{i_1 \dots l \dots i_n;j_1 \dots j_n}(z_1 \dots z_n;z_1' \dots z_n')\nonumber\\
 &= \pm\i\hbar\sum_{lmn}\int_{\mathcal{C}}\d\bar{z}\;\Big\{ W_{i_klmn}(z_k\bar{z}) \,
 \hat{G}^{(n+1)}_{i_1 \dots m \dots i_nn;j_1 \dots j_nl}(z_1 \dots z_n\bar{z};z_1' \dots z_n'\bar{z}^+)\Big\}\nonumber\\
 &\quad -\sum_{p=1}^n \Big\{(\pm1)^{k+p}\delta_{i_kj_p}\delta_{\mathrm{C}}(z_kz_p') \,
\hat{G}^{(n-1)}_{i_1...\bcancel{i_k}...i_n;j_1...\bcancel{j_p}...j_n}(z_1...\bcancel{z_k}...z_n;z_1'...\bcancel{z_p'}...z_n')\Big\}\;.\label{eq:corr_tr}
\end{align}
The question remains whether these reversed equations have a solution and what the relation between this solution and the solution of the original  (non-reversed) equations is. 
Applying $\hat{T}$ from the left and $\hat{T}^{-1}$ from the right on both sides of Eq. \eqref{eq:corr_element}, and omitting the time arguments for brevity, yields
\begin{align}
 &\sum_l\left[-\i\hbar\dfrac{\partial}{\partial z_k}\delta_{i_kl}-h_{i_kl}^{(0)}(z_k)\right]\hat{T}\hat{G}^{(n)}_{i_1...l...i_n;j_1...j_n}\hat{T}^{-1}\nonumber\\
 &= \mp\i\hbar\sum_{lmn}\int_{\mathcal{C}}\d\bar{z}\; W_{i_klmn}(z_k\bar{z})\hat{T}\hat{G}^{(n+1)}_{i_1...m...i_nn;j_1...j_nl}\hat{T}^{-1}\nonumber\\
 &\quad +\sum_{p=1}^n (\pm1)^{k+p}\delta_{i_kj_p}\delta_{\mathrm{C}}(z_kz_p')\hat{T}\hat{G}^{(n-1)}_{i_1...\bcancel{i_k}...i_n;j_1...\bcancel{j_p}...j_n}\hat{T}^{-1}\;.\label{eq:TcorrT}
\end{align}
This is not equivalent to Eq.~\eqref{eq:corr_tr}, therefore $\hat{T}\hat{G}^{(n)}_{i_1...l...i_n;j_1...j_n}\hat{T}^{-1}$ does not solve the reversed equations. 

We, therefore, use a different approach which takes advantage of the fact that $\hat{G}^{(n)}_{i_1 \dots l \dots i_n;j_1 \dots j_n}$ can be interpreted as a functional of $\hat{c}_{i_1}$, \dots , $\hat{c}_{i_n}$, $\hat{c}^{\dagger}_{j_1}$, \dots , $\hat{c}^{\dagger}_{j_n}$. Considering that $\hat{T}\hat{c}_{i}\hat{T}^{-1}$ and $\hat{T}\hat{c}^{\dagger}_{j}\hat{T}^{-1}$ solve the reversed equations of motion compared to $\hat{c}_{i}$ and $\hat{c}^{\dagger}_{j}$, it could be expected that the solution to the reversed hierarchy equations is given by the same functional $\hat{G}^{(n)}_{i_1...l...i_n;j_1...j_n}$ of $\hat{T}\hat{c}_{i}\hat{T}^{-1}$ etc. This is, in fact, the case because
\begin{align}
 &\hat{T}\hat{G}^{(n)}_{i_1 \dots l \dots i_n;j_1 \dots j_n}\hat{T}^{-1}\nonumber\\
 =& \;\hat{T}\left\{ \left(-\frac{\i}{\hbar}\right)^n\hat{\mathcal{T}}_{\mathcal{C}} \hat{c}_{i_1}(z_1) \dots
 \hat{c}_{i_n}(z_n)\hat{c}^{\dagger}_{j_n}(z_n')\dots \hat{c}^{\dagger}_{j_1}(z_1')\right\}\hat{T}^{-1}
 \nonumber\\ =&
 \;(-1)^n \left(-\frac{\i}{\hbar}\right)^n\Big\{\hat{\mathcal{T}}_{\mathcal{C}} \hat{T}\hat{c}_{i_1}(z_1)\hat{T}^{-1}\dots \hat{T}\hat{c}_{i_n}(z_n)\hat{T}^{-1}
\hat{T}\hat{c}^{\dagger}_{j_n}(z_n')\hat{T}^{-1}\dots \hat{T}\hat{c}^{\dagger}_{j_1}(z_1')\hat{T}^{-1}\Big\}\nonumber\\
 =& \;(-1)^n\hat{G}^{(n)}_{i_1\dots l\dots i_n;j_1\dots j_n}\Big\{ \hat{T}\hat{c}_{i_1}(z_1)\hat{T}^{-1}\dots \hat{T}\hat{c}_{i_n}(z_n)\hat{T}^{-1}
 \hat{T}\hat{c}^{\dagger}_{j_n}(z_n')\hat{T}^{-1}...\hat{T}\hat{c}^{\dagger}_{j_1}(z_1')\hat{T}^{-1}\Big\}\nonumber\\
 =:&\; (-1)^n \tilde{\hat{G}}^{(n)}_{i_1...l...i_n;j_1...j_n}\;.
\end{align}
Inserting this into Eq.~\eqref{eq:TcorrT} yields
\begin{align}
 &(-1)^n\sum_l\left[-\i\hbar\dfrac{\partial}{\partial z_k}\delta_{i_kl}-h_{i_kl}^{(0)}(z_k)\right]\tilde{\hat{G}}^{(n)}_{i_1\dots l\dots i_n;j_1\dots j_n}\\
 &= \mp(-1)^{n+1}\i\hbar\sum_{lmn}\int_{\mathcal{C}}\d\bar{z}\; W_{i_klmn}(z_k\bar{z})\tilde{\hat{G}}^{(n+1)}_{i_1\dots m\dots i_nn;j_1\dots j_nl}
 +(-1)^{n-1}\sum_{p=1}^n (\pm1)^{k+p}\delta_{i_kj_p}\delta_{\mathrm{C}}(z_kz_p')\tilde{\hat{G}}^{(n-1)}_{i_1\dots \bcancel{i_k}\dots i_n;j_1\dots \bcancel{j_p}\dots j_n}\;,
 \nonumber
\end{align}
which, when divided by $(-1)^n$, is equivalent to Eq.~\eqref{eq:corr_tr}. 
From this it follows directly, by taking the ensemble average of both sides, that $\mathrm{G}^{(n)}\left[\hat{T}\hat{c}\hat{T}^{-1}, \hat{T}\hat{c}^{\dagger}\hat{T}^{-1}\right]$ satisfies the reversed 
$n^{\mathrm{th}}$-order equations of the Martin--Schwinger hierarchy in the same way. Thus, we have demonstrated that the exact Martin-Schwinger hierarchy is time reversal invariant, as expected.

\newcommand{\TcT}{\left[\hat{T}\hat{c}\hat{T}^{-1}\right]}
\newcommand{\TctT}{\left[\hat{T}\hat{c}^{\dagger}\hat{T}^{-1}\right]}
\section{Time Reversal Invariance of $\Phi$-derivable Approximations}\label{s:tri-phi}
\noindent
Since the solution of the Martin-Schwinger hierarchy is usually possible only with suitable approximations, the important question arises which approximations retain the time reversal properties of the exact system. In the following we demonstrate that any $\Phi$-derivable selfenergy leads to time reversal invariance.
Thereby we will restrict ourselves to real-valued Hamiltonians,  $\hat{H}^*=\hat{H}$.
\subsection{Time Reversal Symmetry Condition for the Selfenergy}
\noindent
Let us recall the first Kadanoff--Baym equation,
\begin{align}
 &\left[ \i\hbar\partial_z - h(z)\right] G_{[\hat{c}]}(zz') 
 = \delta_{\mathcal{C}}(zz')\mathbb{1} + \int_{\mathcal{C}}\d\bar{z}\,\Sigma_{[G_{[\hat{c}]}]}(z\bar{z})G_{[\hat{c}]}(\bar{z}z')\;,
\end{align}
and take the complex conjugate of both sides,
\begin{align}
 &\left[ -\i\hbar\partial_z - h(z)\right] G^*_{[\hat{c}]}(zz') 
 = \delta_{\mathcal{C}}(zz')\mathbb{1} + \int_{\mathcal{C}}\d\bar{z}\,\Sigma_{[G_{[\hat{c}]}]}^*(z\bar{z})G_{[\hat{c}]}^*(\bar{z}z')\;,
\end{align}
where
$ G^*_{[\hat{c}]}(zz') = -G_{[\hat{c}^*]}(zz') $
and, therefore,
\begin{align}
 &-\left[ -\i\hbar\partial_z - h(z)\right] G_{[\hat{c}^*]}(zz')
 = \delta_{\mathcal{C}}(zz')\mathbb{1} - \int_{\mathcal{C}}\d\bar{z}\,\Sigma_{[G_{[\hat{c}]}]}^*(z\bar{z})G_{[\hat{c}^*]}(\bar{z}z')\;. 
\end{align}
This means that $G_{[\hat{c}^*]}$ solves the reversed equation
\begin{align}
 &\left[ -\i\hbar\partial_z - h(z)\right] G_{[\hat{c}^*]}(zz') = -\delta_{\mathcal{C}}(zz')\mathbb{1} 
 -\int_{\mathcal{C}}\d\bar{z}\,\Sigma_{[G_{[\hat{c}^*]}]}^{z^{(\prime)}\rightarrow-z^{(\prime)}}(z\bar{z})G_{[\hat{c}^*]}(\bar{z}z')\;,
\end{align}
if the following holds true for the selfenergy $\Sigma$:
\begin{equation}
\label{eq:selfenergy_condition}
 \Sigma^*_{[G_{[\hat{c}]}]} = -\Sigma^{z^{(\prime)}\rightarrow-z^{(\prime)}}_{[G_{[\hat{c}^*]}]}\;,
\end{equation}
where the superscript denotes that the sign of both $z$ and $z'$ is inverted.
\subsection{$\Phi$-derived selfenergies}
\noindent
Consider the important case of selfenergies that are expressed as a functional derivative of a scalar potential $\Phi$. 
Complex conjugation of both sides of Eq. \eqref{eq:sigma_phi} yields
\begin{equation}
 \Sigma^* = \dfrac{\delta\Phi^*[G]}{\delta G^*} = - \dfrac{\delta\Phi^*[G_{[\hat{c}]}]}{\delta G_{[\hat{c}^*]}}\;,
\end{equation}
and, therefore, condition \eqref{eq:selfenergy_condition} for the selfenergy translates into the following condition for the functional $\Phi$:
\begin{equation}
 \Phi^*[G_{[\hat{c}]}] = \Phi^{z^{(\prime)}\rightarrow-z^{(\prime)}}[G_{[\hat{c}^*]}]\;.
\end{equation}
The rules governing the construction of valid functionals $\Phi$ dictate [\onlinecite{stefanucci}] that a $n^{\mathrm{th}}$-order diagram includes $2n$ contour-time integrals, $2n$ single-particle Green functions $G$, $n$ interparticle interactions $W$ and a factor $(\i\hbar)^n$. 
This means that
\begin{align}
 \Phi^*[G_{[\hat{c}]}] &= (-1)^n \Phi[G^*_{[\hat{c}]}] =(-1)^n \Phi[-G_{[\hat{c}^*]}]\nonumber\\
 &= (-1)^{3n} \Phi[G_{[\hat{c}^*]}] = \Phi^{z^{(\prime)}\rightarrow-z^{(\prime)}}[G_{[\hat{c}^*]}]\;.
 \label{eq:phi_tri}
\end{align}
The last equivalence is true because of the delta-functions in the $n$ interparticle interactions and the $2n$ contour-time integrals that lead to $3n$ sign changes under time-reversal. Thus we have shown that any $\Phi$-derivable NEGF approximation is time reversal invariant.

\subsection{Example: Hartree--Fock selfenergy}
\noindent
The simplest example of a $\Phi$-derivable selfenergy is Hartree-Fock. Nevertheless, it is instructive to explicitly verify that $\Sigma^{\mathrm{HF}}$ satisfies equation \eqref{eq:selfenergy_condition}. To this end we take the complex conjugate of both sides of equation \eqref{eq:sigma_hf}:
\begin{align}
 \Sigma^{\mathrm{HF},*}(zz') &= \pm(-\i)\hbar \delta_{\mathcal{C}}(zz')\int_{\mathcal{C}}\d\bar{z}\,W(z\bar{z})\left[-\mathrm{G}_{[\hat{c}^*]}^{(1)}(\bar{z}\bar{z}^+)\right]\nonumber
 + (-\i)\hbar \left[-\mathrm{G}_{[\hat{c}^*]}^{(1)}(zz')\right]\,W(z^+z')\nonumber\\
 &= \pm\i\hbar \delta_{\mathcal{C}}(zz')\int_{\mathcal{C}}\d\bar{z}\,W(z\bar{z})\,\mathrm{G}_{[\hat{c}^*]}^{(1)}(\bar{z}\bar{z}^+) \nonumber
 + \i\hbar \mathrm{G}_{[\hat{c}^*]}^{(1)}(zz')\,W(z^+z')\nonumber\\
  &= \pm\i\hbar \left[-\tilde{\delta}_{\mathcal{C}}(zz')\right]\int_{\mathcal{C}}\d(-\bar{z})\,\left[-\tilde{W}(z\bar{z})\right]\,\mathrm{G}_{[\hat{c}^*]}^{(1)}(\bar{z}\bar{z}^+) \nonumber
 + \i\hbar \mathrm{G}_{[\hat{c}^*]}^{(1)}(zz')\left[-\tilde{W}(z^+z')\right]\nonumber\\
 &= -\Sigma^{\mathrm{HF},\,z^{(\prime)}\rightarrow-z^{(\prime)}}_{[G_{[\hat{c}^*]}]}\;,
\end{align}
where $\tilde{\delta} = \delta^{z^{(\prime)}\rightarrow-z^{(\prime)}}$ and $\tilde{W} = W^{z^{(\prime)}\rightarrow-z^{(\prime)}}$. 
Equivalently, it can be checked that $\Phi^{\mathrm{HF}}$ satisfies Eq.~\eqref{eq:phi_tri}:
\begin{align}
 \Phi^{\mathrm{HF},*} =
 & \pm\dfrac{(-\i)}{2} \int_{\mathcal{C}}\int_{\mathcal{C}}\d z_1\d z_2\;\left[-G_{[\hat{c}^*]}(z_1;z_1^+)\,W(z_1;z_2)\right]\left[-G_{[\hat{c}^*]}(z_2;z_2^+)\right] \nonumber\\
 &\quad + \dfrac{(-\i)}{2}\int_{\mathcal{C}}\int_{\mathcal{C}}\d z_1\d z_2\;\left[-G_{[\hat{c}^*]}(z_1;z_2^+)\right]W(z_1;z_2)\left[-G_{[\hat{c}^*]}(z_2;z_1^+)\right]\nonumber\\
 &= \mp\dfrac{\i}{2} \int_{\mathcal{C}}\int_{\mathcal{C}}\d z_1\d z_2\;G_{[\hat{c}^*]}(z_1;z_1^+)\,W(z_1;z_2)\,G_{[\hat{c}^*]}(z_2;z_2^+) \nonumber\\
 &\quad  - \dfrac{\i}{2}\int_{\mathcal{C}}\int_{\mathcal{C}}\d z_1\d z_2\;G_{[\hat{c}^*]}(z_1;z_2^+)\,W(z_1;z_2)\,G_{[\hat{c}^*]}(z_2;z_1^+)\nonumber\\
 &= \pm\dfrac{\i}{2} \int_{\mathcal{C}}\int_{\mathcal{C}}\d(-z_1)\d(-z_2)\;G_{[\hat{c}^*]}(z_1;z_1^+)\left[-\tilde{W}(z_1;z_2)\right]\,G_{[\hat{c}^*]}(z_2;z_2^+) \nonumber\\
 &\quad + \dfrac{\i}{2}\int_{\mathcal{C}}\int_{\mathcal{C}}\d(-z_1)\d(-z_2)\;G_{[\hat{c}^*]}(z_1;z_2^+)\left[-\tilde{W}(z_1;z_2)\right]\,G_{[\hat{c}^*]}(z_2;z_1^+)\nonumber\\
 &=\Phi^{\mathrm{HF},\,z^{(\prime)}\rightarrow-z^{(\prime)}}[G_{[\hat{c}^*]}]\;.
\end{align}

\section{Summary and Discussion}\label{s:summary}
In this paper, it has been explicitly shown that the governing equations of the nonequilibrium Green functions formalism, the exact Martin--Schwinger hierarchy and the associated quantum-kinetic equations are time reversible. This is in striking contrast to conventional Boltzmann-type kinetic equations where irreversibility is introduced by the ``Stoßzahlansatz'' or similar procedures. The existence of generalized quantum kinetic equations that retain the reversibility of the underlying quantum-mechanical equations is known for a long time. Here we have presented a simple procedure that allows to verify this property. It is based on use of the anti-unitary time-reversal operator ${\hat T}$ that translates the solution of the Schrödinger equation into the time-reversed equation.

We then turned to approximate solutions  to the NEGF formalism that are based on approximations of the selfenergy. We have demonstrated that any selfenergy that is $\Phi$-derivable is symmetric with respect to time reversal, as long as the (single particle) Hamiltonian possesses an anti-unitary symmetry $\hat{H} = \hat{T}\hat{H}\hat{T}^{-1}$. 
These approximations include the well-known Hartree--Fock, second Born and T-Matrix approximations as well as many others. 

Aside from the $\Phi$-derivable selfenergy approximations discussed above, in recent years another class of approximations has attracted high interest: the generalized Kadanoff-Baym ansatz (GKBA). It replaces the two-time Green function by a  single-time approximation. The GKBA was originally derived by Lipavsky {\em et al.} [\onlinecite{lipavsky}], and a rigorous derivation from density operator theory was given in Ref.~[\onlinecite{bonitz_qkt}].
In a detailed investigation by Hermanns {\em et al.} [\onlinecite{hermanns_prb_14}] it was shown that the GKBA retains the conservation properties of the original two-time equations if the approximation for the retarded Green function $G^R$ is conserving as well. The same reasoning can be applied to the issue of time reversal invariance. The result is that use of a $\Phi$-derivable approximation for $G^R$ (which may differ from the approximation for the selfenergy) will retain the time reversal properties of the original two-time approximation.

An interesting outcome of our analysis is that $\Phi$-derivable  approximations for the selfenergy are both conserving and time reversible. It remains to investigate whether this applies also to other classes of approximations. Finally, proof of time-reversibility of an approximation is also of high practical value in numerical solutions of the KBE as this provides a sensitive test for the numerical accuracy and convergence, e.g.~[\onlinecite{stan-comment}].


\end{document}